\documentclass[12pt]{article}

\newcommand{\de}{\partial}
\newcommand{\be}{\begin{equation}}
\newcommand{\ee}{\end{equation}}
\newcommand{\bea}{\begin{eqnarray}}
\newcommand{\eea}{\end{eqnarray}}
\newcommand{\bd}{\begin{displaymath}}
\newcommand{\ed}{\end{displaymath}}
\newcommand{\nn}{\nonumber\\}

\def\NPB{{\em Nucl. Phys. }}
\def\PLB{{\em Phys. Lett. }}  

\def\PRD{{\em Phys. Rev. }} 
 
\def\MPLA{{\em Mod. Phys. Lett. }}

\def\lb{\label}
\def\to{\rightarrow}

\def\lq{
\left[}
\def\lt{\left(}

\def\rq{\right]}
\def\rt{\right)}

\def\g{\gamma}

\def\e{\eta}

\def\m{\mu}

\def\l{\lambda}
\def\r{\rho}
\def\s{\sigma}

\def\t{\tau}
\def\lp{L_{P}}

\begin{document}
\thispagestyle{empty}
\vskip 1.5cm
\begin{flushright}
 INFNCA-TH0005  
\end{flushright}
\vskip 3.5cm
\centerline{\large \bf Thermodynamical Behaviour of Composite} 
\centerline{\large \bf Stringy Black Holes}
\vspace{2.2cm}
\centerline{\bf M. Cadoni and C.N. Colacino}
\vskip 0.5cm
\centerline{\sl Dipartimento di Fisica, 
Universit\`a degli studi di Cagliari}  
\centerline{\sl I.N.F.N. \ - \  Sezione di Cagliari, }
\centerline{\sl S.P. Monserrato-Sestu, km 0,700 I-09042 Monserrato, 
Italy}
\vskip 3.2cm
\centerline{\large \bf ABSTRACT}
{We study the thermodynamical and geometrical behaviour of the black 
holes 
that arise as solutions of the heterotic string action. We discuss 
the 
near-horizon  scaling behaviour of the solutions that are described by 
two-di\-men\-sio\-nal Anti-de Sitter space (AdS$_{2}$). We find that 
finite-energy 
excitations of  AdS$_{2}$ are suppressed only for scaling limits 
characterized 
by a dilaton   constant near the horizon, whereas this suppression does 
not 
occur when the dilaton is  non-constant. }
\vskip 0.3cm
\hrule
\begin{flushleft}
{E-Mail: CADONI@CA.INFN.IT\hfill}
\end{flushleft}
\begin{flushleft}
{E-Mail:COLACINO@CA.INFN.IT\hfill}
\end{flushleft}
\newpage
\section{Introduction}
The charged black hole solutions of string theory in four dimensions 
\cite{vari, behrndt,youm} have been investigated for different purposes. 
From the point of view of General Relativity (GR) they generalize the 
Reissner-Nordstrom (RN) solutions of the Einstein-Maxwell  theory. 
From the 
point of view of string theory they represent low-energy geometrical 
structures that should give some information about the fundamental 
string 
dynamics \cite {DR,PT}.

More recently, these solutions have become interesting also from the 
point view of the Anti-de Sitter (AdS)/Conformal Field Theory (CFT) 
duality \cite{mama,ooguri}. In fact, it is well known that in the 
near-horizon, near-extremal regime the RN-like charged black hole 
solutions of 
string theory behave like $AdS_{2}\times S^{2}$. They can be, 
therefore, very useful for trying to understand better the puzzles of 
the  AdS$_{2}$/CFT$_{1}$ duality \cite{stromi, cami}.
In Ref. \cite{mamistro} a detailed study was performed about the 
arising of
$AdS_{2}$ as near-horizon geometry of the RN solution of the 
Einstein-Maxwell 
theory. It was found that the finite-energy excitations of $AdS_{2}$
are suppressed. Only zero-energy configuration survive, whose 
degeneration should, in principle, be able to explain the entropy of 
the near-extremal RN black hole. 

In this paper we extend the discussion of Ref. \cite{mamistro} 
to the black 
hole solutions in the context of heterotic 
string theory. Owing to the string/string /string triality of 
heterotic, type IIA and type IIB strings in four dimensions 
\cite{DLR}, our discussion also holds for the black hole solutions of 
type IIA and type IIB string theory.
It is well known that the most general solution of this kind 
represents a 
generalization of the RN black hole.
The new feature with respect to the 
RN case is represented by the presence of scalar fields (the dilaton 
and the moduli).

Using duality symmetry arguments we argue that the relevant 
information about the solutions is encoded in a single-scalar 
single-U(1)-field  solution.
We analyze in detail both from the geometrical and 
thermodynamical point of view this black hole solution.
We find that, whereas the geometrical structure of the 
the near-horizon solutions is the same as in the pure RN case, the 
presence of the dilaton allows both for solutions with constant 
dilaton and solutions with  non-constant dilaton. 
Whereas in the former case finite-energy excitations of $AdS_{2}$ 
are still suppressed in the latter they are allowed.

The structure of the paper is the following. In Sect. 2 we show as 
the general black hole solution of heterotic string theory can be 
written as a single-scalar single-U(1) field solutions with RN causal 
structure. In Sect. 3 we discuss the geometrical and the thermodynamical 
behaviour of the solutions in the near-extremal, near-horizon limit.

\section{Black hole solutions }
The truncated version of the bosonic action for the heterotic string 
compactified on a six-torus is the following \cite{maharana}
\bea
S&=& \frac{1}{16\pi G}\int d^{4}x\sqrt{g}
\Big\{R-\frac{1}{2}\left[ (\de\eta)^{2}+
(\de\tau)^2+(\de\rho)^2\right]\nonumber\\
\mbox{}& &-\frac{e^{-\eta}}{4}\left[e^{-\tau-\rho}F_1^2+
e^{-\tau+\rho}F_2^2+e^{\tau+\rho}F_3^2+e^{\tau
-\rho}F_4^2\right]
\Big\}.
\label{prima}\eea

In this action, we have set to zero the axion fields and all the 
$U(1)$ fields
but four, two Kaluza-Klein fields $F_1$, $F_2$ and two winding modes 
$F_3$, $F_4$. The scalar fields are related to the standard 
definitions of the string coupling, 
K\"{a}hler form and complex 
structure of the torus,
\be
e^{-\eta}=Im S\qquad e^{-\tau}=Im T\qquad e^{-\rho}=Im U.
\ee

The most general, non-extremal (dyonic) solution in the Einstein-Hilbert
frame is given by \cite{CY,behrndt}
\bea
ds^2&=&-\lt H_1H_2H_3H_4\rt^{1/2}fdt^2+\lt H_1H_2H_3H_4\rt^{-1/2}
\lt f^{-1}dr^2+r^2d\Omega_2^2\rt,\nn
\mbox{}e^{2\eta}&=&\frac{H_1H_3}{H_2H_4}, \quad 
e^{2\tau}=\frac{H_1H_4}{H_2H_3}, \quad 
e^{2\rho}=\frac{H_1H_2}{H_3H_4}\nn
F_1&=&dH_1\wedge dt, \quad \tilde F_2=dH_2\wedge dt,
\quad F_3=dH
_3\wedge dt, \quad \tilde F_4=dH_4\wedge dt
\label{soluzioni}
\eea
where  $\tilde{F}_{2}= 
 {e^{-\e-\tau+\r}}{^*}F_{2},\,\tilde{F}_{4}= {e^{-\e+\tau-\r}}{^*}F_{4}$
 ($^{*}$ denotes the Hodge dual) and $H_i, f$, $i=1\ldots 4$, are
  given in terms of  harmonic functions, 
\be
H_i=\lt g_i+\frac{\mu\sinh^2\alpha_i}{r}\rt^{-1},
\quad f=1-\frac{\mu}{r}.\label{silvia}
\ee
The extremal limit is obtained by 
\be
\mu\to0\qquad\sinh^2\alpha_{i}¥\to\infty\qquad\mu\sinh^2\alpha_{i}¥\to 
q_{i}.
\label{extremal}\ee
For particular values of the parameters the solutions 
(\ref{soluzioni}) can be 
put in correspondence with the solutions of the effective dilaton 
gravity 
action
\be
S_{eff}=\frac{1}{16\pi G}\int d^4 x\sqrt{g}\lq R-2\lt\de\Phi\rt^2
+e^{-2a\Phi}F^2\rq\label{efficace}
\ee
with  $a$ given by one of the following
four values \cite{behrndt}
\be
a=0\qquad a=\frac{1}{\sqrt3}\qquad a=1\qquad a=\sqrt3
\label{e1}\ee
The case $a=0$ describes  the Reissner-Nordstr\"{o}m black hole of GR 
and corresponds to $H_{1}= H_{2}=H_{3}=H_{4}$ in Eq. 
(\ref{soluzioni}), whereas $a=\sqrt3, \,a=1\,a=1/\sqrt3$ 
correspond, respectively, to $H_{2}=H_{3}=H_{4}=1$, 
$ H_{1}=H_{2},\, H_{3}=H_{4}=1$, $H_{1}=H_{2}=H_{3},\, H_{4}=1$.

A very interesting proposal is the so called {\it compositeness 
idea}, 
according to which the $a=\sqrt3$ solution can be seen as a 
fundamental state
of which the other solutions are bound states with zero binding energy
\cite{DR,rahm,cadoni}. This idea stems basically from the higher 
dimensional 
interpretation of black holes as intersections of D-branes. The number of 
individual components is denoted by $n$.
Elementary $n=1$
solutions correspond to dilaton gravity theories with 
$a=\sqrt3$;
$n=2$ bound states correspond to $a=1$ solutions, $n=3$ to 
$a=1/\sqrt3$ 
and finally $n=4$ correspond to $a=0$. Moreover, the compositeness 
idea has been used together with the duality symmetries (in 
particular 
the $O(3,Z)$ duality group) of the model 
to generate the whole spectrum of BPS states \cite{cadoni}.

The purpose of this paper is to study in detail the thermodynamical and 
geometrical behaviour of the stringy composite black holes 
(\ref{soluzioni}) and  to establish which 
relation they have towards  Reissner-Nordstr\"{o}m black holes.

The general solution (\ref{soluzioni}) is very complicated to study.
It contains nine arbitrary integration constants  (four moduli 
$g_{i}$, four $U(1)$-charges and the mass); thanks to the $O(3,Z)$ 
duality symmetry of the model, all the relevant
information about the nature of these black holes can be 
obtained by studying some 
simplified models which we get as we move in the moduli and in the 
charge 
space.  Following the spirit of the compositeness idea, 
we can study those solutions we get if we equate some of the charges 
and of the moduli. 
In this way we will construct solutions with $\alpha_{i}\neq 0$
that describe single-scalar 
single-U(1)-field  black holes and  whose strong (or weak) coupling 
regime
is exactly given by the dilaton gravity solutions of the model  
(\ref{efficace}).
  
This can be done in a systematic 
way by exploiting the $O(3,Z)$ duality 
symmetry of the model in a way similar to that followed in 
Ref. \cite{cadoni}
in dealing with the case of some null charges.
The solutions can be characterized by 
giving the number $m$ of equal moduli and charges (we consider only 
solutions with the same number of equal charges and moduli).
It is evident that $m$ is invariant under the action of the $O(3,Z)$ 
duality group described in Ref. \cite{cadoni}. It can be therefore 
used to label different representations of the duality group. Because 
$m=1$ is equivalent to $m=3$ we will have  three multiplets on which 
the duality group  $O(3,Z)$ will act by changing the scalar and the 
$U(1)$-field but leaving the geometry of the solution unchanged.

Hence, it will be enough to consider just one representative solution 
for each multiplet, the whole multiplet can be obtained acting on 
this 
solution with the $O(3,Z)$ group. 
Because $m=4$ is nothing but the 
well-known  Reissner-Nordstrom solution, in the following we will 
consider 
only $m=N=1,2,3$. We will set  $\alpha_{1}=\alpha_{2}=\alpha_{4}$,  
$g_{1}=g_{2}=
g_{4}$ for $N=1,3$ and $\alpha_{1}=\alpha_{3}$, 
$\alpha_{2}=\alpha_{4}$, $g_{1}=g_{3}$, $g_{2}=g_{4}$ for $N=2$.
Requiring the solution  to be asymptotically 
Minkowskian will impose an additional constraint on  the moduli:
$\prod_{i=1}^{4}g_{i}=1$.

The  solution  can be written in a simple form by
introducing the parameters $\lambda_{i}$
\be
\lambda_{i}=\frac{\mu\sinh^2\alpha_i}{g_i}\nonumber
\lb{e2},\ee
the scalar
charge $\sigma$ and the parameters $r_{\pm}$, defined 
as follows:
\bea
r_{-}&=&\l_{1},\quad r_{+}=\mu+\l_{1}, \quad 2 \s 
L_{P}=\l_{3}-\l_{1},\quad{\rm for} N=1,\nn
r_{-}&=&\l_{1},\quad r_{+}=\mu+\l_{1}, \quad 2 \s 
L_{P}=\l_{2}-\l_{1}, \quad{\rm for} N=2,\nn
r_{-}&=&\l_{3},\quad r_{+}=\mu+\l_{3}, \quad 2 \s 
L_{P}=\l_{1}-\l_{3}, \quad{\rm for} N=3,
\lb{e3}\eea
where $L_{P}$ is the Planck length. The $U(1)$-charges $q_i$ can be written
in terms of the other parameters as follows (no summation on i)
\be
q_i^2=g_i^2\lambda_i\left(\lambda_i+\m\right)\label{panunzio}
\ee

With the previous positions and performing the coordinate change
$r\to r- r_{-}$ the solution (\ref{soluzioni}) becomes:

\bea\label{ngenerale}
ds^2&=&-\frac{(r-r_{+})(r-r_{-})}{r^{(4-N)/2}(r+2\sigma 
L_P)^{N/2}}dt^2
+\frac{r^{(4-N)/2}(r+2\sigma 
L_P)^{N/2}}{(r-r_{+})(r-r_{-})}dr^2\nonumber\\
\mbox{}&+&r^{(4-N)/2}(r+2\sigma 
L_P)^{N/2}d\Omega_2^2\nn
e^{\e}&=&e^{\e_0}\lt1+\frac{2\s\lp}{r}\rt^{\g},
\eea

where $\g=1$ for $N=2$ and $\g=\pm1/2$  for $N=1,3$ respectively.
For $N=1,3$ we also have $\t=\r=-\e$ while for $N=2$ we have $\t=\r=0$ 
as can be easily seen from eq. (\ref{soluzioni}).
The duality group $O(3,Z)$ acting  on the solutions  
(\ref{ngenerale}) 
with a given  $N$ generates the  corresponding multiplet. The 
duality acts on the scalars whereas the 
metric part of the solution remains unchanged. For instance, the $\tau_{S}$ 
duality \cite{cadoni} ($\e\to -\e, F_{1}\to\tilde F_{3}, F_{3}\to\tilde F_{1}, 
F_{2}\to\tilde F_{4}, F_{4}\to\tilde F_{2}$), acting 
on the solution $N=2$, exchanges the electric solutions with the magnetic ones.

The parameters $r_{+},r_{-},\s$ are related to the mass $M$ and 
U(1)-charges $q_i$ by the following 
equations
\be 
\lp^{2} Q^{2}= r_{+}r_{-},\quad 2M \lp^{2}= r_{+}+r_{-}+ N\s\lp,
\lb{h1}\ee
where we have introduced the adimensional charge $Q$ defined as
\be
Q\equiv\frac{q_i}{L_Pg_i}\label{panelli}
\ee
where $i=1$ for $N=1,2$ and $i=3$ for $N=3$.
From eq. (\ref{h1}) follows the relation
\be
r_{\pm}= \lp \left(\lp M-{N\over2}\s\pm\sqrt{(\lp M 
-{N\over2}\s)^{2}-Q^{2}}\right)
\lb{h2}\ee
as well as the extremality condition
\be
\lp M\ge Q+{N\over2} \s.
\lb{h3}\ee

The solutions (\ref{ngenerale}) represent a four-parameters 
generalization of the RN solution of GR.
In the heterotic string context, the RN solution is described by the 
$m=4$ multiplet, and it is obtained by putting $N=0$ in
Eq. (\ref{ngenerale}). 

One can easily show that the causal structure of the solutions 
(\ref{ngenerale}) is the 
same as that of the RN solutions, the effect of the scalar charge 
$\sigma$ being a shift of the $r=0$ singularity. 
Solutions of this kind have been already discussed in the literature 
\cite{ cagib}.

It is interesting to notice that the bound-state solutions with 
$n=1,2,3$ elementary constituents can be obtained as strong (or weak 
for the dual solutions) 
coupling regime of the corresponding $m=1,2,3$ solution. In fact, the 
former solutions are characterized by some vanishing $U(1)
$-charges, 
whereas from Eq. (\ref{e2}) and (\ref{extremal}) it follows that 
$q=0$ can be achieved by 
$g\to\infty$. This fact has a natural explanation if one uses the 
bound state interpretation of the black holes. Since the bound 
state has zero binding energy, the mass of the composite state is the sum 
of the masses $M_{i}$ of the elementary constituents. 
But $M_{i}$ behaves as ${q_{i}/g_{i}}$ so that in the strong 
coupling regime some of the elementary constituents of the $n=4$ 
solution do not contribute, leaving an effective  $n=1,2,3$ bound 
state.
   
\section{Thermodynamical behaviour}

An important property of the RN black holes is  that the 
semiclassical 
analysis of
their thermodynamical behaviour breaks down very near extremality 
\cite{preskill}: the formulae for the entropy and temperature are 
given by
\bea
S_{BH}&=&\frac{\pi r_{+}^2}{L^2_P}\nonumber\\
T_{H}&=&\frac{1}{4\pi}\left(\frac{\de g_{00}}{\de r}\right)_{r=r_{+}}
=\frac{r_{+}-r_{-}}{4\pi r^2_{+}}.\label{formulae}
\eea
Near extremality, the excitation energy above extremality $E$ and 
temperature $T_{H}$
are related in the following way
\be
E\equiv 2\pi^{2} Q^3T_{H}^2L_P\label{rigore}. 
\ee
At an excitation energy of
\be
E_{gap}=\frac{1}{Q^3L_P}\label{gap}
\ee
the semiclassical analysis ceases to be valid. The nature of this 
breakdown is
well understood in string theory: the black hole develops a mass gap 
\cite{gap}
and (\ref{gap}) is the energy of its lowest-lying excitation. 
Maldacena,
Michelson and Strominger \cite{mamistro} studied both the geometry and 
the thermodynamical behaviour of the near-extremal RN black 
holes in the near-horizon limit. They showed that the spacetime always 
factorizes as $AdS_2\times S^2$, with $AdS_2$ endowed with the 
Robinson-Bertotti metric.
Moreover, they found that in this limit all the finite-energy 
excitations of $AdS_2$ are suppressed.

The previous features hold for the RN solution. One would like to 
know if they represent a general feature of the solutions with 
$AdS_2\times S^2$ near-horizon geometry. Let us therefore 
consider the near-extremal, near-horizon thermodynamical 
behaviour of the  black hole solutions (\ref{ngenerale}).

The Hawking temperature of these solutions 
is given by
\be
T_{H}=\frac{r_{+}-r_{-}}{4\pi r_{+}^2}\left(\frac{r_{+}+2\sigma L_P}
{r_{+}}\right)^{-N/2},\label{effervescente}
\ee
whereas for the entropy we have,
\be
S=\frac{\pi r^2_{+}}{L_P^2}\left(\frac{r_{+}+2\sigma L_P}{r_{+}}\right)^{N/2}.
\label{aspirina}\ee
As we are going to see see in detail soon, in the near-extremal, 
near-horizon regime the solutions (\ref{ngenerale}) always behave as 
$AdS_2\times S^2$.  
In this situation the  energy-temperature relation is 
\be
E\equiv 2\pi^{2} 
T_H^2L_PQ^3\left(\frac{Q+2\sigma}{Q}\right)^N\label{inzaghi}
\ee
and the energy (\ref{gap}) at which the semiclassical analysis 
breaks down 
is given by
\be
E_{gap}=\frac{1}{Q^3L_P}\left(\frac{Q}{Q+2\sigma}\right)^N.
\label {gaggio}\ee

Eqs. (\ref{inzaghi}) and (\ref{gaggio}) represent a generalization 
to the heterotic string of the formulae (\ref{rigore}) and 
(\ref{gap}) of the pure RN case. 
The only difference between the two cases is the presence of the 
scalar charge $\s$, which is a consequence of the presence of a
non-trivial dilaton.  The near-horizon limit can be achieved as in 
Ref.\cite{mamistro} by letting $\lp\to 0$ but holding fixed some of the 
remaining parameters  $E,T_{H}, Q, \s$. In principle one could also consider 
more general limits for which  $T_{H}$  is not fixed but, as shown
in Ref. \cite{mamistro}, these cases present intricate features, in 
particular the geometry becomes singular.
We performed a detailed analysis of the various limits for the three 
cases ($N=1,2,3$) and found out that for finite $\sigma$ the results obtained 
by Maldacena and others \cite{mamistro} for the RN black hole hold true also 
for these other solutions: the near-horizon geometry is always
$AdS_2\times S^2$  and there are no finite-energy 
excitations, despite the presence of the scalar charge. This result 
was somehow expected but never proved explicitly in literature.

The new feature appears when we take the limit $\lp\to 0$ 
together with $\s\to\infty$, holding $E,T_{H}$ and $Q$ fixed.
In this case the near-horizon geometry is again $AdS_2\times S^2$
and finite energy excitation of $AdS_2$ are allowed holding $\lp 
\s^{N}$ fixed. However in this case $AdS_2$ is endowed with a 
non-constant, linearly varying, dilaton. The resulting 
two-dimensional 
model is similar to that  corresponding to the $a=1/\sqrt 3$ case in Eq. 
(\ref{efficace}), which has been shown to be also relevant for the 
 $AdS_2/CFT_{1}$ correspondence \cite{cami, cadoni1}.
We do not discuss here the limit 
$L_{p}\to 0$, $Q\to\infty$, $(T_{H},\s)$ fixed because there is 
nothing new 
with respect to the pure RN case discussed in Ref. \cite{mamistro}.
The relevant limits to be discussed are:
(a)  $L_{p}\to 0$, $(T_{H},Q,\s)$  fixed; (b)  $L_{p}\to 0$, 
$\s\to\infty$,
$(T_{H},Q)$ fixed. In both cases the near-horizon geometry is 
$AdS_2\times S^2$  but whereas in case (a) the dilaton is constant 
near the horizon in case (b) we have a non-constant, linearly varying dilaton.

{\leftline{(a) $L_{p}\to 0$, $(T_{H},Q,\s)$ fixed }
Defining 
\be
U={r-r_{+}\over \lp^{2}},
\ee
and performing the limit in Eq. (\ref{ngenerale}) keeping $U$ fixed,
 we get
\bea
{ds^{2}\over Lp^{2}}&=& - {U^{2}+4\pi R^{2}T_{H}U\over R^{2}}dt^{2}+
 {R^{2}\over U^{2}+4\pi R^{2}T_{H}U }dU^{2}+ R^{2}d\Omega_{2}^{2},\nn
 e^{\eta}&=&e^{\eta_{0}}\left({R\over Q}\right)^{4\gamma/N},
 \eea
 where $R$ is given in terms of the two charges $\s,Q$:
 \be
 R= Q\left(1+{2\s\over Q}\right)^{N/4},
 \ee
 and $\gamma$ is given as in Eq. (\ref{ngenerale}).
 Hence, in this case the near-horizon geometry is the same as in the 
 pure RN case, the dilaton being a constant near the horizon. What 
 changes is just the radius $R$ of the transverse two-sphere. In our 
 case it is a function of both the $U(1)$- and scalar charges. 
 From Eqs. (\ref{inzaghi}) and  (\ref{gaggio}) we obtain that in this 
limit the excitation energy goes to zero, whereas $E_{gap}$ diverges.
 Analogously to the pure RN case we cannot have finite energy 
 excitations of $AdS_2$.

{\leftline{(b) $L_{p}\to 0$, $\s\to\infty$,
$(T_{H},Q, E)$ fixed} 
 
From Eqs.  (\ref{inzaghi}) and  (\ref{gaggio}) it is 
evident 
that we can hold  
$(T_{H},Q, E)$ fixed while $L_{p}\to 0$ if we allow $\s\to\infty$, with
$L_p\s^{N}\equiv const.$   
Let us define 
\be
V=\left ({E_{gap}\over \lp^{3}Q}\right)^{1/2}(r-r_{+}).
\ee
Performing the limit while keeping $V$ fixed, the solution  
(\ref{ngenerale})
becomes,
\be
{ds^{2}E_{gap}\over Lp}= - (V^{2}+4\pi T_{H}V) dt^{2}+
 {1\over V^{2}+4\pi T_{H}V }dV^{2}+ d\Omega_{2}^{2},\quad
 {\eta}\propto \ln{V}.
\ee

As in the previous case the near-horizon geometry is $AdS_2\times 
S^2$, but now the dilaton is not constant. 
Thus,  heterotic string black holes allow for finite-energy 
excitations of $AdS_2$, but they require a non constant dilaton.
This feature has been also noticed in Ref. \cite{mamistro} 
in the analysis of the pure RN case. 

It is interesting to notice that $AdS_2$ with a nonconstant 
dilaton has already emerged as the near-horizon geometry of the $n=3$ 
($a=1/\sqrt{3}$ in Eq. (\ref{efficace})) heterotic string black 
hole \cite{cadoni1}. $AdS_2$  endowed with a nonconstant 
dilaton has been also used to give a realization of the 
$AdS_2/CFT_{1}$ correspondence. Here we have shown that this kind of models 
can emerge also as near-horizon limit of RN-like four dimensional geometries.
Moreover, the fact that in this context finite energy excitations of 
$AdS_2$ are allowed could be useful to circumvent some of the problems
of the pure RN case.  
 
Until now we have considered only solutions with all the four 
$U(1)$-charges 
different from zero. Let us now consider  the solutions we obtain 
when some of the charges go to zero, i.e the composite black hole 
solutions with $n=1,2,3$. Although the thermodynamics of these black 
holes has been already discussed in the literature \cite{HW}, 
it is interesting 
to compare it with the thermodynamical behaviour of the $N=1,2,3$ 
solutions.

The solutions describing bound states of $n=1,2,3$ elementary black 
holes can be obtained as the $g\to \infty$, strong coupling regime, of 
the  solution with $N=1,2,3$. Because of Eq. (\ref{e2}) the 
corresponding solutions can be obtained putting $r_{-}=0$ in  
Eq.(\ref{ngenerale}). The inner horizon disappears and the causal 
structure of the  solutions becomes radically different from that of 
the RN-like black holes. In the extremal limit the horizon matches 
the 
singularity which is timelike for $n=1$ and null for $n=2,3$.

The Hawking temperature associated with the black hole is 
\be
T=\frac{1}{4\pi}\left(\frac{\de g_{00}}{\de 
r}\right)_{r=r_{+}}=\frac{1}
{4\pi}\left(r_{+}\right)^{{n\over 2}-1}\left(r_{+}+2\sigma 
L_P\right)^{-n/2}.
\label{nakata}
\ee

In the near-extremal limit $r_{+}=2EL_P^2$ and thus we have 
the following energy-temperature relations
\bea 
E&\sim&\left(64\pi^2L_P^3\sigma T^2\right)^{-1}, \quad {\rm for}\quad 
n=1,\nn
T&\sim&\frac{1}{8\pi\sigma L_P}, \quad {\rm for}\quad n=2,\nn
E&\sim& 64\pi^2T^2L_P\sigma^3 , \quad {\rm for}\quad n=3.
\lb {p2}\eea
For $n=1$ the specific heat is negative and this is probably related 
with the nature of the singularity, which in the extremal limit is 
timelike. For $n=2$ there is no dependence of the excitation 
energy on the temperature. This behaviour can be explained, at least 
in principle, in terms of the underlying two-dimensional model 
\cite{cadoni1,donami}.
Finally, for  $n=3$, the energy-temperature relation is similar to 
that   of Eq. (\ref{rigore}). This relation indicates that model has 
a sensible description in terms of an effective two-dimensional model 
that admits $AdS_2$ as solution \cite{cadoni1}. The main difference with the 
RN-like case is that here the dilaton is not constant near the horizon.

\end{document}